\documentclass[a4paper,12pt]{article}
\usepackage{amssymb}
\input{epsf}
\newcommand{\spacing}[1]{\renewcommand{\baselinestretch}{#1}\large\normalsize}
\spacing{1}
\makeatletter
\def\@maketitle{%
  \newpage\spacing{1}\setlength{\parskip}{12pt}%
    {\Large\bfseries\noindent\sloppy \textsf{\@title} \par}%
    {\noindent\sloppy \@author}%
}
\makeatother

\makeatletter
\renewcommand*{\@biblabel}[1]{\hfill#1.}
\makeatother

\makeatletter
\def\@cite#1#2{$^{\mbox{\scriptsize #1\if@tempswa , #2\fi}}$}
\makeatother

\newenvironment{affiliations}{%
    \setcounter{enumi}{1}%
    \setlength{\parindent}{0in}%
    \slshape\sloppy%
    \begin{list}{\upshape$^{\arabic{enumi}}$}{%
        \usecounter{enumi}%
        \setlength{\leftmargin}{0in}%
        \setlength{\topsep}{0in}%
        \setlength{\labelsep}{0in}%
        \setlength{\labelwidth}{0in}%
        \setlength{\listparindent}{0in}%
        \setlength{\itemsep}{0ex}%
        \setlength{\parsep}{0in}%
        }
    }{\end{list}\par\vspace{12pt}}

\title{Gravitational redshift of galaxies in clusters as predicted by general relativity}
\author{Rados{\l}aw Wojtak$^{1}$, Steen H. Hansen$^{1}$ \& Jens Hjorth$^{1}$}

\begin{document}

\maketitle

\begin{affiliations}
\item Dark Cosmology Centre, Niels Bohr Institute, University of Copenhagen, Juliane Maries Vej 30, DK-2100 Copenhagen \O, 
Denmark
\end{affiliations}

\noindent

\textbf{The theoretical framework of cosmology is mainly defined by gravity, of 
which general relativity is the current model. Recent tests of 
general relativity within the $\Lambda$ Cold Dark Matter (CDM) model have found 
a concordance between predictions and the observations of the growth 
rate and clustering of the cosmic web\cite{Rap10,Rey10}. General relativity has 
not hitherto been tested on cosmological scales independent of the 
assumptions of the  $\Lambda$CDM model. Here we report observation of the 
gravitational redshift of light coming from galaxies in clusters at 
the $99$ per cent confidence level, based upon archival data\cite{Aba09}. The 
measurement agrees with the predictions of general relativity and its 
modification created to explain cosmic acceleration without the need 
for dark energy ($f(R)$ theory\cite{Car04}), but is inconsistent with 
alternative models designed to avoid the presence of dark matter\cite{Mil83,Bek04}.}

According to the theory of general relativity\cite{Ein16}, light emitted from galaxies moving 
in the gravitational potential well of galaxy clusters is expected to be redshifted proportionally 
to the difference in gravitational potential $\Phi$ between the clusters and an observer, i.e., 
$z_{\rm gr}=\Delta\Phi/c^{2}$, where $c$ is the velocity of light in vacuum. For typical cluster masses of $\sim 10^{14}M_{\odot}$, 
where $M_{\odot}$ is the Sun's mass, the gravitational 
redshift is estimated to be\cite{Cap95,Bro00,Kim04} $cz_{\rm gr}\approx 10$ km s$^{-1}$ which is around two orders 
magnitude smaller than the Doppler shift due to the random motions of galaxies in clusters. The method of disentangling 
the kinematic Doppler effect from gravitational redshift relies on the fact that the former gives rise to a symmetric 
broadening of the observed velocity distribution, whereas the latter shifts its centroid. A critical factor in detecting 
such a velocity shift is the number of galaxies with spectroscopically measured velocities and the number 
of galaxy clusters. Both should be sufficiently high in order to reduce the error due to the Doppler width of the velocity 
distribution and eliminate the sensitivity to irregularities in cluster structure, e.g. substructures, asphericity. 

The data are compiled from the SDSS\cite{Aba09} Data Release 7 and the associated  Gaussian Mixture Brightest Cluster Galaxy 
catalogue\cite{Hao10} containing the positions and redshifts of galaxy clusters 
identified in the survey. The cluster sample is richness-limited with a threshold corresponding to a cluster mass of 
$10^{14}M_{\odot}$. The mean, 5- and 95-percentile values of the cluster richness\cite{Hao10} 
are $16$, $8$, and $86$ and correspond to cluster masses of around $2\times 10^{14}M_{\odot}$, $10^{14}M_{\odot}$ and 
$10^{15}M_{\odot}$. The typical number of spectroscopic redshifts per cluster (within a $6$ Mpc aperture and a $\pm 4000$ km s$^{-1}$ 
velocity range around the mean cluster velocities) varies from $10$ for low-richness clusters to $140$ for the richest ones.

Fig.~1 shows the histograms of galaxy velocities calculated in four bins of the projected cluster-centric distance 
centred at $0.6$, $1.6$, $3.3$ and $5.2$ Mpc. The cluster centres and redshifts were approximated by the coordinates and 
redshifts of the brightest cluster galaxies, hereafter BCGs. The observed velocity distributions consist of two clearly 
distinct parts: a quasi-flat distribution of galaxies not belonging to the clusters (observed due to projection effect) 
and a quasi-Gaussian component associated with galaxies gravitationally bound to the clusters\cite{Woj07}. The latter is 
expected to reveal the signature of gravitational redshift in terms of a systematic shift of its velocity centroid. Analysis 
of mock kinematic data generated from cosmological simulations shows that the number of redshifts and clusters is sufficient 
to reduce all expected sources of noise such as substructures, cluster asphericity, non-negligible off-set between BCGs and 
clusters centres\cite{Ski11} (both in the position on the sky and redshift space), and to allow for detection of gravitational 
redshift at nearly $3\sigma$ confidence level (see SI).

We search for gravitational redshift by measuring the mean velocity $\Delta$ of the quasi-Gaussian component of the 
observed velocity distribution. We carry out a Monte Carlo Markov Chain analysis of the data using a 
two-component model for the velocity distribution which includes both a contribution from the cluster and non-cluster 
galaxies (SI). Constraints on the mean velocity are obtained by marginalising the likelihood function 
over the set of nuisance parameters defining the shape of both components of the velocity distribution. 
The best fitting models of the velocity distributions are shown in Fig.~1 and the resulting measurements of the 
mean velocity as a function of the projected cluster-centric distance $R$ are presented in Fig.~2. The obtained 
mean velocity is negative at all radii with a clear tendency to 
decline with increasing radius. The negative values arise from the fact that the rest frames of the clusters 
are defined by the observed velocities of the central galaxies. This choice of the reference frame implies 
that the gravitational redshift manifests itself as a blueshift\cite{Kim04} (negative mean velocity) varying 
with the projected cluster-centric distance from $0$ at the cluster centre to $-|\Phi(0)|/c$ at large projected 
radii $R$.

The detection of gravitational redshift is significant at the $99$ per cent confidence level. The integrated signal within the 
$6$ Mpc aperture amounts to $\Delta=-7.7\pm3.0$ km s$^{-1}$ which is consistent with the gravitational potential 
depths of simulated galaxy clusters of\cite{Kim04} $\Delta=-(5-10)$ km s$^{-1}$. A more quantitative comparison 
with theoretical predictions requires explicit information about the mean gravitational potential profile and 
the distribution of cluster masses in the sample. We make use of the velocity dispersion profile of the composite 
cluster to constrain both functions. Then we calculate the gravitational redshift in terms of the mean velocity 
$\Delta$ by convolving the individual profiles of the clusters with their mass distribution (SI). 
The resulting profile (red profile in Fig.~2; see also discussion on the effect of the anisotropy of galaxy orbits in SI) 
is fully consistent with the gravitational redshift inferred from the velocity distributions. The fact that the same 
gravitational potential underlies galaxy motions and gravitational redshift of photons in clusters provides observational 
evidence of the equivalence principle on the scale of galaxy clusters.

We confront the obtained constraints on gravitational redshift with the predictions of alternative theories of 
gravity. We consider two popular models of gravity, the tensor-vector-scalar (hereafter TeVeS) theory\cite{Mil83,Bek04} 
and the $f(R)$ model\cite{Car04}, designed to alleviate the problem of dark matter or to recover the expansion history of 
the Universe, respectively. Theoretical profiles of gravitational redshift are calculated using the relations between 
the generalised gravitational potentials of these models and the Newtonian potential (SI). The Newtonian potential is inferred from the 
observed velocity dispersion profile of the composite cluster under the assumption of the most reliable anisotropic model 
of galaxy orbits (see SI for more details), and constitutes the reference basis for the calculations. For TeVeS 
we assume that the total masses of galaxy clusters make up $80$ per cent of those recovered under assumption of the Newtonian 
gravity. This factor lowers the ratio of dynamical-to-baryonic mass in galaxy clusters to the value resulting from 
fitting Modified Newtonian Dynamics\cite{Mil83} (to which TeVeS is a relativistic generalisation) to cluster 
data\cite{Poi05}. The resulting profile of gravitational redshift does not match the data, deviating from the observations at the 
$95$ percent confidence level (the blue dashed line in Fig.~3). This discrepancy increases with projected radius 
and is mostly caused by a logarithmic divergence of the scalar field in the regime of small accelerations, i.e., 
$g<a_{0}$ and $a_{0}\approx 10^{-10}$ m s$^{-2}$, which is responsible for a $1/r$ modification of the gravitational acceleration. 
This result points to a critical problem for TeVeS (or Modified Newtonian Dynamics) in recovering the 
true gravitational potential at large distances around the cluster centres. Considering the $f(R)$ model, we choose 
the least favourable set of free parameters maximising the departure from Newtonian gravitational acceleration\cite{Sch10}. 
Despite this choice, the resulting profile of gravitational redshift is consistent with the data (the blue solid line in 
Fig.~2).

The obtained constraints on gravity are consistent with recent tests verifying the concordance between gravity, 
cosmological model and observations of the large scale structure of the Universe\cite{Rap10,Rey10}. An important 
advantage of using gravitational redshift effect is that this method does not depend on cosmology (see also SI) 
allowing to probe gravity in a direct way. In particular, this implies that the discrepancy between TeVeS theory 
and the observations\cite{Rey10} is unlikely to be a consequence of a specific 
choice of cosmological parameters, but indeed points to the inadequacy of this model to describe the Universe on 
very large scales.

Our results complement a series of experiments and observations aimed at confirmation of the predicted gravitational 
redshift on different scales of the Universe. Fig.~3 shows a summary of the detections in terms of the 
relative accuracy of the measurements as a function of the scale of the gravitational potential well. The positions 
of data points vary from $20$ m, for the first ground-based experiment\cite{Pou59,Pou64}, to the $1-10$ Mpc scale for 
galaxy clusters. On the scale of the solar radius we plot the measurement of gravitational redshift for the Sun\cite{Lop91}, 
and on the $2$ orders of magnitude smaller scale constraints from the observations of Sirius B white dwarf\cite{Gre71,Bar05} 
and space-borne hydrogen maser\cite{Ves80}. These results make gravitational redshift the only effect predicted by general 
relativity which has been confirmed on spatial scales spanning $22$ orders of magnitude. Studying this effect in more detail 
relies on the size of the redshift sample and therefore will be possible with the advent of the next generation redshifts 
surveys, e.g., the EUCLID satellite.

\newpage

\section*{Acknowledgments}
The Dark Cosmology Centre is funded by the Danish National Research Foundation. R.W. wishes to thank D. Rapetti, G. Mamon 
and S. Gottl\"ober for fruitful discussions and suggestions. The mock catalogues of galaxy clusters have been obtained 
from a simulation performed at the Altix of the LRZ Garching.

\section*{Author contributions}
R.W., analysis of the velocity distributions and the velocity dispersion profile, predictions for the models of modified gravity, drafting the manuscript; 
S.H.H., comparison with the constraints on gravitational redshift on different scales, writing and commenting the paper; 
J.H., conceiving the idea of the measurement, writing and commenting the paper;

\newpage

\begin{center}
\leavevmode
\epsfxsize=16cm
\epsfbox[50 50 580 420]{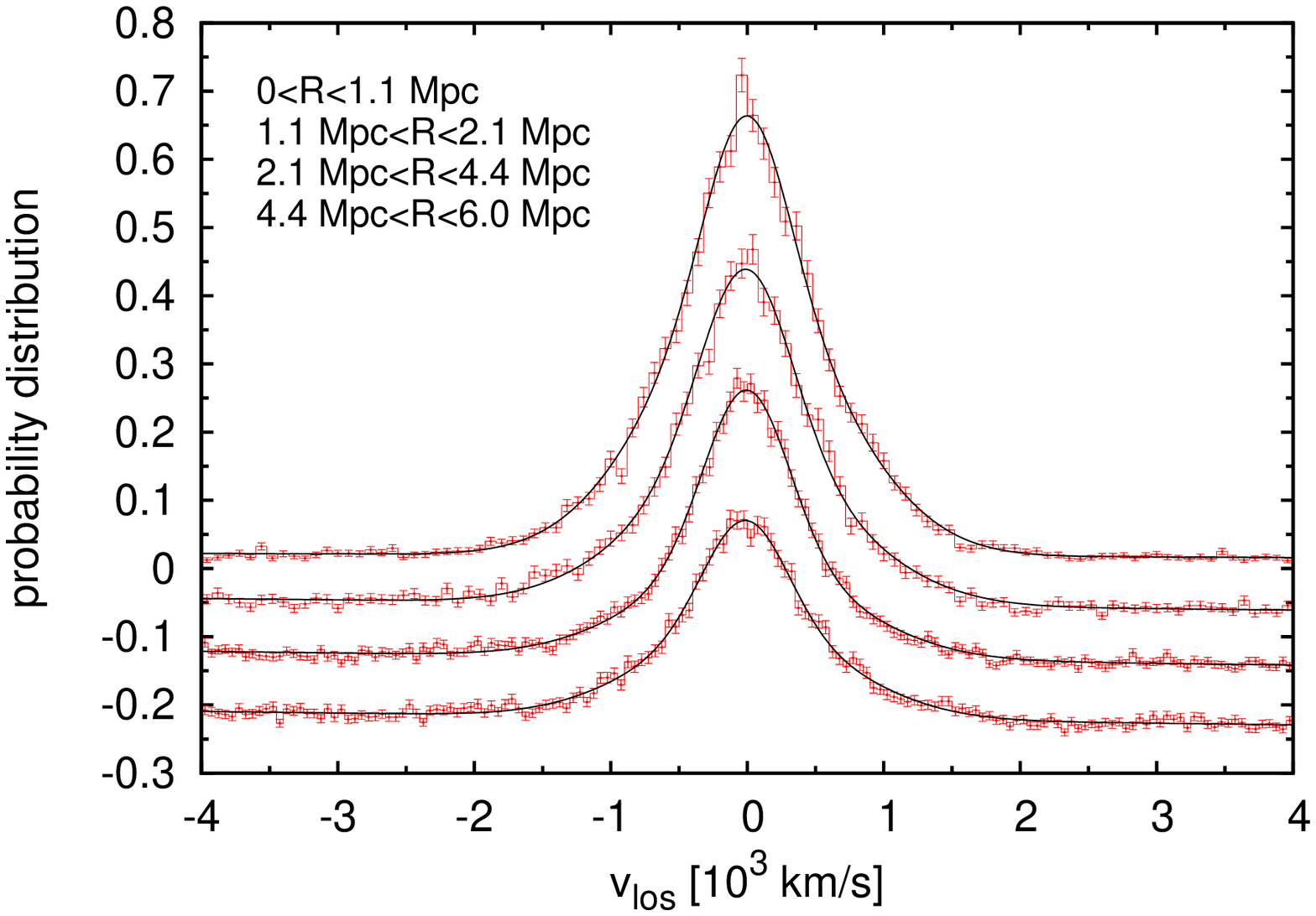}
\end{center}
{\sf \textbf{Figure 1} Velocity distributions of galaxies combined from $7,800$ SDSS galaxy clusters. 
The line-of-sight velocity ($v_{\rm los}$) distributions are plotted in four bins of the projected cluster-centric distances $R$. 
They are sorted from the top to bottom according to the order of radial bins indicated 
in the upper left corner and offset vertically by an arbitrary amount for presentation purposes. 
Red lines present the histograms of the observed galaxy velocities in the cluster rest frame and black solid 
lines show the best fitting models. The model assumes a linear contribution from the galaxies which do not 
belong to the cluster and a quasi-Gaussian contribution from the cluster members (see SI for more details). 
The cluster rest frames and centres are defined by the redshifts and the positions of the brightest cluster galaxies. 
The error bars represent Poisson noise.}

\newpage
\begin{center}
\leavevmode
\epsfxsize=16cm
\epsfbox[50 50 580 420]{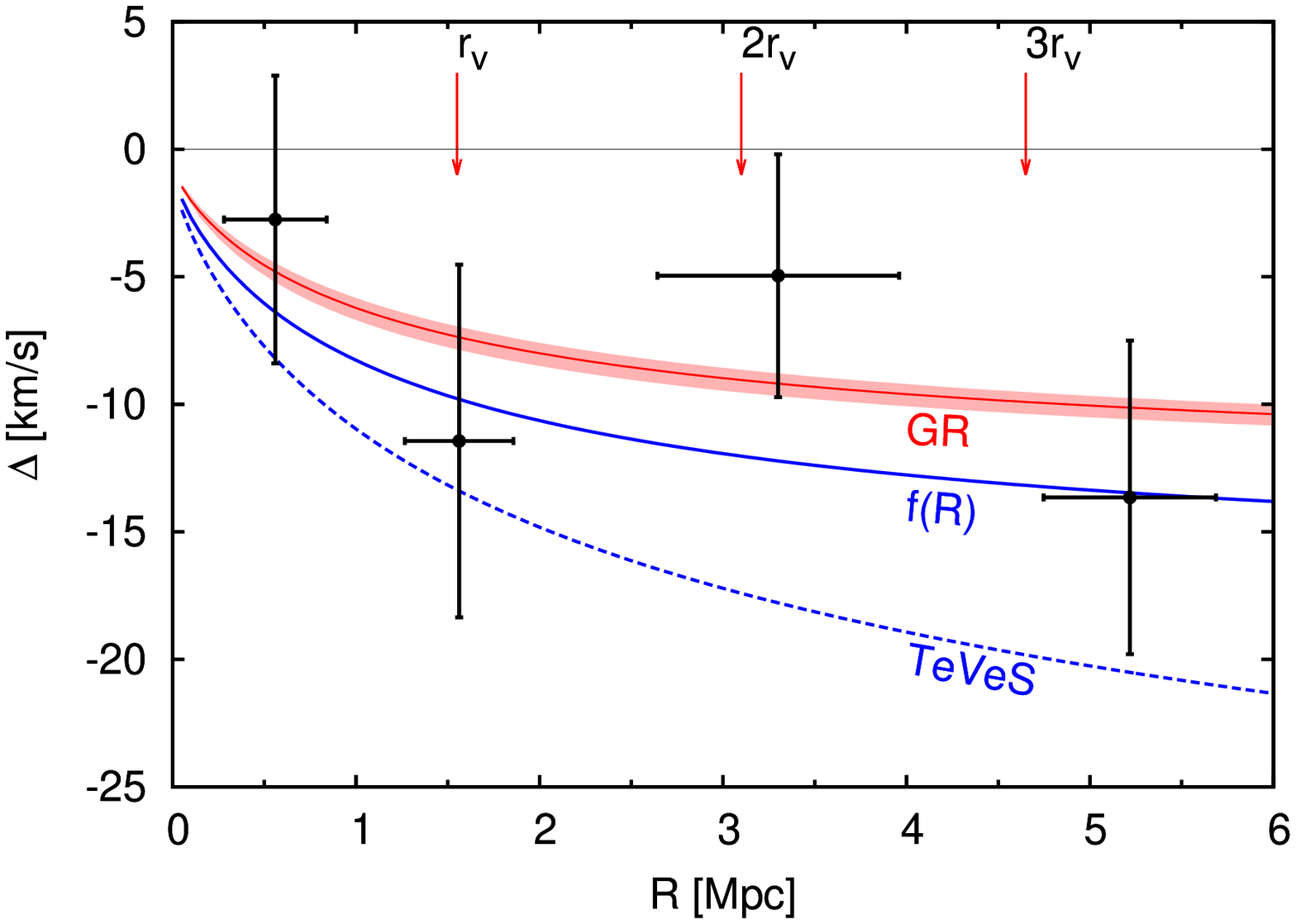}
\end{center}
{\sf \textbf{Figure 2} Constraints on gravitational redshift in galaxy clusters. The effect manifests itself 
as a blueshift $\Delta$ of the velocity distributions of cluster galaxies in the rest frame of their BCGs. Velocity 
shifts were estimated as the mean velocity of a quasi-Gaussian component of the observed velocity distributions 
(see Fig.~1). The error bars represent the range of $\Delta$ parameter containing $68$ per cent of the 
marginal probability and the dispersion of the projected radii in a given bin. The blueshift (black points) 
varies with the projected radius $R$ and its value at large radii indicates the mean gravitational potential 
depth in galaxy clusters. The red profile represents theoretical predictions of general relativity calculated on 
the basis of the mean cluster gravitational potential inferred from fitting the velocity dispersion profile under 
the assumption of the most reliable anisotropic model of galaxy orbits (see SI for more details). Its width 
shows the range of $\Delta$ containing $68$ per cent of the marginal probability. The blue solid 
and dashed lines show the profiles corresponding to two modifications of standard gravity: $f(R)$ theory\cite{Car04} 
and the tensor-vector-scalar (TeVeS) model\cite{Mil83,Bek04}. Both profiles were calculated on the basis of the 
corresponding modified gravitational potentials (see SI for more details). The prediction for $f(R)$ represents the case 
which maximises the deviation from the gravitational acceleration in standard gravity on the scales of galaxy clusters. 
Assuming isotropic orbits in fitting the velocity dispersion profile lowers the mean gravitational depth of the 
clusters by $20$ per cent. The resulting profiles of gravitational redshift for general relativity and $f(R)$ theory 
are still consistent with the data and the discrepancy between prediction of TeVeS and the measurements remains nearly 
the same. The arrows show characteristic scales related to the mean radius $r_{\rm v}$ of the virialized parts of 
the clusters.}

\newpage
\begin{center}
\leavevmode
\epsfxsize=16cm
\epsfbox[50 50 580 420]{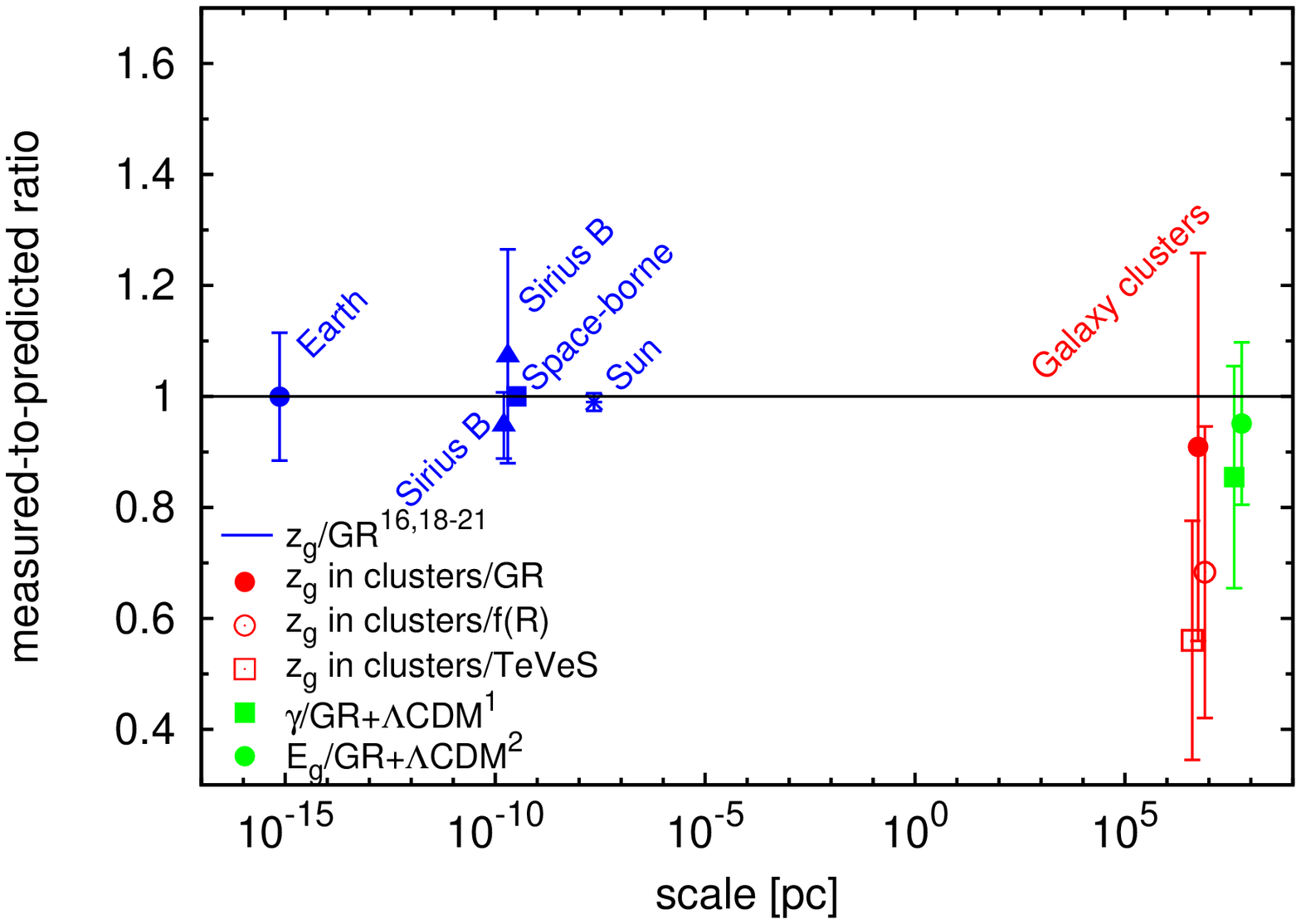}
\end{center}
{\sf \textbf{Figure 3} The measured-to-predicted ratio of the gravitational redshift. 
The figure shows the results of different observations or experiments 
as a function of the spatial scale of the gravitational potential well. Blue and red symbols refer to detections of gravitational redshift $z_{\rm g}$ in: ground-based experiment\cite{Pou59} (blue circle), observations of Sirius B white dwarf\cite{Gre71,Bar05} (blue triangles), space-based experiment\cite{Ves80} (blue square), observation of the Sun\cite{Lop91} (blue star), analysis of the cluster data reported in this work (red circle). All measurements are compared with the 
predictions of general relativity (solid symbols). Results obtained for galaxy clusters are also compared with the 
predictions of $f(R)$ theory and TeVeS model (red empty symbols). As a measure of gravitational redshift in galaxy clusters 
we used the signal integrated within the aperture of $6$ Mpc. The green square and circle show the measurement of the rate of 
growth of cosmic structure\cite{Rap10} and the probe of gravity $E_{\rm g}$ combining the properties of galaxy-galaxy 
lensing, galaxy clustering and galaxy velocities\cite{Rey10}. Both results are compared with the prediction of general relativity with a 
standard $\Lambda$CDM cosmological model. All error bars represent standard deviations. The relative accuracy of the 
measurement from space-born experiment\cite{Ves80} is beyond the resolution of the plot and amounts to $10^{-4}$.}

\newpage

\begin{center}
{\Large {\sf \textbf{Supplementary Information}}}
\end{center}

\section*{{\large Measuring gravitational redshift of galaxies in clusters}}

The gravitational potential depth in typical galaxy clusters, expressed in 
terms of the velocity shift, is estimated at around\cite{Kim04} $-10$ km s$^{-1}$. This is two orders of magnitude 
smaller than the Doppler shift arising from random motions of galaxies in clusters. 
Bearing in mind that the signature of the gravitational redshift lies in the mean of the velocity distribution one can 
show that in order to reduce the error on the gravitational redshift to the level of the effect itself one 
needs at least $10^{4}$ velocities. This number should grow by a factor of $4$, if one requires a $2\sigma$ 
detection of the effect, and probably by another factor of $2$ in order to account for a non-negligible 
number of background galaxies which do not contribute to the effect, but give rise to the shape of the observed 
velocity distribution. Needles to say, the only means to collect as many as 
$10^{4}-10^{5}$ galaxy velocities in clusters is stacking redshift data of sufficiently large number of clusters 
($>10^{3}$ clusters, assuming that current redshift surveys typically provide $10$ redshifts per cluster). 

Combining redshift data from a number of clusters also allows to reduce the error 
resulting from local irregularities of velocity distributions in individual clusters. Such irregularities 
arise naturally from the presence of substructures or filaments along the line of sight, deviation from 
spherical symmetry, residual streaming motions etc. In order to address the impact of these factors on 
the error of the gravitational redshift estimate, one needs to refer to cosmological simulations. Such 
an analysis was carried out by Kim \& Croft\cite{Kim04} who concluded that the minimum number 
of galaxy clusters required to confirm the gravitational redshift effect at the $2\sigma$ confidence level 
is $\sim 3000$. With this number of clusters the gravitational redshift may be traced up to $6$ Mpc, which is $3-4$ 
times larger than the size of the virialized part of clusters-- a natural boundary condition for 
all methods of the mass measurement based on the assumption of virial equilibrium.

Another important source of inaccuracy in the measurement of gravitational redshift, which was partly 
taken into account by Kim \& Croft\cite{Kim04}, is the choice of a central cluster galaxy which ideally would be an 
object at rest at the bottom of the gravitational potential well. In our work, we approximate 
such galaxy by a brightest cluster galaxy, hereafter BCG. In general, such choice is not fully justified 
because the positions and velocities of BCGs exhibit some deviations from those defined by the cluster 
mass centres\cite{Lin07,Ski11}. For example, the typical dispersion of the random velocities of BCGs may reach 
$30-40$ per cent of the total velocity dispersion in galaxy clusters\cite{Ski11}. Yet, among all cluster 
galaxies, BCGs are those whose positions and velocities coincide mostly with the location and velocities of 
the cluster centres. In order to reduce the error caused by the non-vanishing velocities of BCGs to the 
level required for detection of the gravitational redshift effect, one needs to combine the data from a 
sufficiently large number of clusters, e.g.,  around $2500$ clusters for a $2\sigma$ detection\cite{Kim04}.

\section*{{\large Data}}
In order to compile statistically uniform and possibly the largest sample of galaxy redshifts in clusters, 
we make use of the SDSS\cite{Aba09} Data Release 7 whose integral part is a flux-limited spectroscopic 
survey providing the redshifts of nearly million galaxies brighter than Petrosian $r$-magnitude $17.77$ over 
the area $7400$ deg$^2$. The positions and redshifts of galaxy clusters come from a Gaussian Mixture 
Brightest Cluster Galaxy cluster catalogue\cite{Hao10} which is the most up-to-date catalogue of galaxy 
clusters assembled on the basis of the SDSS DR7. The catalogue comprises $55,000$ galaxy clusters at redshifts 
$0.1<z<0.55$ detected by means of searching for red-sequence galaxies and BCGs. It provides positions and redshifts 
of BCGs residing in galaxy clusters selected from the SDSS above a certain richness limit. For the purpose of our 
analysis, we neglect all clusters whose BCGs do not have spectroscopic redshifts. This reduces the cluster sample 
by $63$ per cent without affecting the relative fractions of poor and rich clusters. 

We approximate the coordinates and redshift of cluster centres by the positions and redshifts of BCGs. Then 
we search for all galaxies within a $6$ Mpc aperture around the cluster centres. The radius of this aperture 
is $\approx 3.5$ larger then the virial radius and corresponds to the turn-around radius\cite{Cup08} at which the expansion 
of the Universe starts to dominate over peculiar velocities of galaxies and the velocity cut-off separates all 
potential cluster galaxies from the galaxies of background or foreground. In order to separate potential 
cluster members from distant interlopers (galaxies of background or foreground), we select only those galaxies 
whose velocities $v_{\rm los}$ in the rest frame of a related BCG, i.e.
\begin{equation}
v_{\rm los}=c\frac{z-z_{\rm BCG}}{1+z_{\rm BCG}},
\end{equation}
where $z$ and $z_{\rm BCG}$ are the redshifts of a given galaxy and related BCG respectively, lie within the 
$\pm 4000$ km s$^{-1}$ range. This velocity cut-off is sufficiently wide to include all cluster members with no 
respect to the cluster mass (the minimum velocity cut-off corresponding to the most massive clusters at small radii 
is around $\pm 3000$ km s$^{-1}$; however, wider velocity range is required for precise modelling of the interloper 
contribution to the observed velocity distribution\cite{Rin03}).

Our final sample comprises $7,800$ clusters with the mean redshift of $0.24$ and on average $16$ galaxies with 
spectroscopic redshift per cluster. Those clusters with less than $5$ redshifts were not included into the 
sample. As the final step we combine redshift data of all clusters into one. The velocity diagram of the 
resulting composite cluster is shown in Supplementary Fig.~1.

\section*{{\large Analysis of the velocity distributions}}

To place constraints on the mean of the observed velocity distribution of cluster galaxies 
induced by the gravitational redshift effect we carry out a Monte Carlo Markov Chain (MCMC) 
analysis of the velocity distribution with the likelihood function defined by
\begin{equation}\label{like}
L=\prod_{i=1}^{N}f(v_{\rm los,i}|\Delta,{\mathbf a}),
\end{equation}
where $f(v_{\rm los}|\Delta,{\mathbf a})$ is a model of the velocity distribution, $\Delta$ is the 
mean of the velocity distribution of cluster galaxies, ${\mathbf a}$ is a vector of nuisance 
parameters describing the shape of the velocity distribution and $N$ is the number of redshifts. 
In order to account for the presence of the interlopers (galaxies of foreground or background 
observed due to projection effect), we use the following two-component model 
of the velocity distribution\cite{Woj07}
\begin{equation}\label{pdf}
f(v_{\rm los})=(1-p_{\rm cl})f_{\rm b}(v_{\rm los}|{\mathbf a})+p_{\rm cl}f_{\rm cl}(v_{\rm los}|\Delta,{\mathbf a}),
\end{equation}
where $f_{\rm cl}(v_{\rm los})$ and $f_{\rm b}(v_{\rm los})$ are the velocity distributions of the cluster 
members and the interlopers, respectively (both normalised to $1$), and $p_{\rm cl}$ is a free 
parameter describing the probability of a given galaxy to be a cluster member. The choice of the functional 
form of $f_{\rm b}(v)$ depends on the operational definition of cluster membership and may vary from a 
wide Gaussian distribution\cite{Mam10}, if one regards all galaxies beyond the virial sphere 
as the background, to a uniform distribution\cite{Woj07}, if only gravitationally unbound 
galaxies contribute to the background. For the purpose of our study a uniform background 
is appropriate, since all gravitationally bound galaxies, regardless of their 
positions with respect to the virial sphere, contribute to the expected signal of gravitational 
redshift. However, close inspection of the data reveals that a uniform model of the background 
must be generalised to account for a subtle asymmetry between the number of interlopers 
with negative and positive velocities. This asymmetry arises from the fact that a flux-limited 
limited survey tends to include slightly more galaxies which are closer and, therefore, have 
negative Hubble velocities in the cluster rest frame. We find that in order to 
account for this effect it is sufficient to assume that $f_{\rm b}(v_{\rm los})$ is linear in 
velocity $v_{\rm los}$. The slope of this velocity distribution is the second nuisance parameter 
of our model, after $p_{\rm cl}$. We also note that the velocity distribution of cluster members cannot 
exhibit a similar asymmetry because it is dominated by random motions of galaxies 
which are independent of the positions with respect to the cluster centre.

The observed velocity distributions of cluster galaxies are not Gaussian at all projected radii $R$ 
(see Fig.~2). This deviation from Gaussianity is expected and arises mostly 
from combining data from clusters of different masses\cite{Dia96}, from the fact that the radial bins are 
wider than the scale of variation of the velocity dispersion profile and from an intrinsic 
non-Gaussianity of velocity distributions of individual clusters\cite{Woj09}. Modelling these 
effects is beyond the scope of this work and, for our purpose, it is sufficient to invoke a 
phenomenological model of $f_{\rm cl}(v_{\rm los})$ providing a satisfactory fit to the data. 
We find that approximating $f_{\rm cl}(v_{\rm los})$ by a sum of two Gaussians with the same 
mean velocity $\Delta$ satisfies this condition. This introduces three additional parameters 
into the model given by eq.~(\ref{pdf}): two velocity dispersions and the ratio of the relative 
weights of both Gaussian components. Performing a K-S test, we verified that the final fits 
of our model are fully consistent with the data ($p=0.99$).

We carry out the MCMC analysis of the velocity distributions in $4$ radial bins of the projected 
cluster-centric distance (see Fig.~2). The number of redshifts in subsequent 
bins varies from $15,000$ in the two innermost bins to $45,000$ for the remaining two. The choice 
of these numbers is motivated by finding a balance between bin spacing and the local number of 
cluster galaxies (proportional to $p_{\rm cl}$ which varies from $0.9$ at the cluster centre 
to $0.3$ at $R\gtrsim 3$ Mpc). BCGs were not included in the first bin, otherwise the estimate 
of the mean would be biased towards $0$. The number of clusters contributing to the subsequent 
bins varies from $\sim 1000$ in the innermost bin to $\sim 2000$ in the outermost one.

For the conversion between the angular and physical scales we adopted a flat $\Lambda$CDM cosmology 
with $\Omega_{\rm m}=0.3$ and the Hubble constant $H_{0}=70$ km s$^{-1}$ Mpc$^{-1}$. We note, 
however, that all galaxy clusters used in this work lie at low redshifts ($z\approx 0.2$) and, 
therefore, the impact of using a particular cosmological model on the final results is negligible.

\section*{{\large Gravitational redshift profile}}

Assuming spherical symmetry, the gravitational redshift profile of a single galaxy cluster 
(in terms of velocity blueshift of the velocity distribution) can be calculated using the 
following formula\cite{Cap95}
\begin{equation}\label{vel-shift-proj}
\Delta_{\rm s} (R)=\frac{2}{c\Sigma(R)}\int_{R}^{\infty}[\Phi(r)-\Phi(0)]\frac{\rho(r)r\textrm{d}r}
{\sqrt{r^2-R^2}},
\end{equation}
where $R$ is the projected cluster-centric distance, $\Phi(r)$ is the gravitational potential, 
$\rho(r)$ and $\Sigma(R)$ are the 3D and surface (2D) density profiles of galaxies. In order to estimate 
this effect for the data combined from a cluster sample, one needs to convolve this expression with 
the distribution of cluster masses in the sample. Then the resulting profile of the blueshift takes 
the following form
\begin{equation}\label{shift-stack}
\Delta (R)=\frac{\int\Delta_{\rm s}(R)\Sigma(R)(\textrm{dN}/\textrm{d}M_{\rm v})\textrm{d}M_{\rm v}}
{\int\Sigma(R)(\textrm{d}N/\textrm{d}M_{\rm v})\textrm{d}M_{\rm v}},
\end{equation}
where $M_{\rm v}$ is the virial mass and $\textrm{d}N/\textrm{d}M_{\rm v}$ is the mass distribution. 
We note that $\Delta_{\rm s}$ depends implicitly on the virial mass and the shape of the gravitational 
potential. The virial mass and radius are defined in terms of the overdensity parameter 
$\delta_{c}=3M_{\rm v}/(4\pi r_{\rm v}^{3}\rho_{c})$, where $\rho_{c}$ is the present critical density. 
In our calculations we adopted $\delta_{c}=102$ (see e.g. {\L}okas \& Hoffman\cite{Lok01}).

The main unknown factor in equation (\ref{shift-stack}) is the mass distribution. 
This may be estimated by means of dynamical modelling of the observed velocity dispersion profile. 
The left panel of Supplementary Fig.~2 shows the velocity dispersion profile estimated in radial bins 
(black points) by fitting eq. (\ref{pdf}) with $f_{\rm cl}(v)$ approximated by a single Gaussian. 
The profile is truncated at $R=1.2$ Mpc which is the virial radius corresponding to the 
anticipated lower limit of all virial masses in the sample which is around $10^{14}M_{\odot}$. We note 
that the velocity dispersion profile is flatter than typical profiles observed in single galaxy 
clusters. This property arises naturally from the fact that the cluster sample is not uniform in terms of 
the cluster mass (more massive clusters give rise to growth of velocity dispersions at large radii). Here 
we use this effect to place constraints on the mass distribution in the cluster sample.

In analogy with equation (\ref{shift-stack}), one can show that the velocity dispersion profile for kinematic 
data combined from a set of clusters can be expressed as
\begin{equation}\label{dispersion}
\sigma_{\rm los}(R)=\Big(\frac{\int\widehat{\sigma}_{\rm los}^{2}(R)\Sigma(R)
(\textrm{dN}/\textrm{d}M_{\rm v})\textrm{d}M_{\rm v}}
{\int\Sigma(R)(\textrm{dN}/\textrm{d}M_{\rm v})\textrm{d}M_{\rm v}}\Big)^{1/2},
\end{equation}
where $\widehat{\sigma}_{\rm los}(r)$ refers to the velocity dispersion profile of a single cluster. We 
approximate the mass distribution by a power-law, i.e., $\textrm{d}N/\textrm{d}M_{\rm v}\propto M_{\rm v}^{-\alpha}$, 
where $\alpha$ is a free parameter. This parameterisation is mostly motivated by the fact that it resembles the 
observed cluster counts as a function of the virial mass\cite{Roz10}. In order to account for 
the richness threshold of the cluster catalogue and cosmological decay of the mass function at high masses we 
impose cut-offs on the mass distribution at low and high masses respectively. The 
cut-offs are fixed at $10^{14}M_{\odot}$ and $2\times10^{15}M_{\odot}$ which are the limits of the 
mass range spanned by the clusters of the maxBCG catalogue\cite{Koe07,Joh07,Roz10}-- the predecessor of 
the Gaussian Mixture Brightest Cluster Galaxy catalogue\cite{Hao10} used in this work.

In order to calculate the velocity dispersion profile of a single cluster, $\widehat{\sigma}_{\rm los}(R)$ in eq. (\ref{dispersion}), 
we make use of a model of the distribution function presented by Wojtak {\it et al.}\cite{Woj08}. The model is constructed 
under assumption of spherical symmetry, constant mass-to-light ratio (galaxies trace dark matter) 
and for a wide range of possible profiles of the orbital anisotropy. We approximate the dark matter density profile 
by the NFW formula\cite{Nav97}, i.e. $\rho(r)\propto 1/[(r/r_{\rm v})(1+c_{\rm v}r_{\rm v})^{2}]$, 
where $c_{\rm v}$ is the concentration parameter. Since fitting the velocity dispersion profile does not allow 
to constrain the mass profile and the anisotropy of galaxy orbits at the same time (the problem known as the mass-anisotropy 
degeneracy), we fix all parameters related to the orbital anisotropy. In order not to loose generality of our analysis 
we consider two models of the orbital anisotropy: anisotropic-- with the anisotropy parameter 
$\beta(r)=1-\sigma_{\theta}^{2}(r)/\sigma_{r}^{2}(r)$ varying from $0$ in the centre to $0.4$ at the virial radius, 
and isotropic with $\beta(r)=0$, where $\sigma_{r}$ and $\sigma_{\theta}$ are the radial and tangential velocity 
dispersions. We note that these two models are two limiting cases of a whole family of the anisotropy profiles found 
both in simulations\cite{Cue08,Han10} and observations\cite{Biv04,Woj10}

We evaluate the velocity dispersion numerically as the second moment of the projected phase-space density at fixed 
projected radii $R$. Then we correct all resulting dispersions for the effect of non-vanishing random velocities of 
BCGs\cite{Ski11}. Such correction relies on replacing all velocity dispersions $\sigma_{\rm los}(R)$ by 
$(\sigma_{\rm los}(R)^{2}+\sigma_{\rm BCG}^{2})^{1/2}$, where $\sigma_{\rm BCG}$ is a typical velocity dispersion 
for BCGs. For our analysis, we assume that $\sigma_{\rm BCG}$ equals $35$ per cent of the total velocity 
dispersion within the virial radius.

Fitting the velocity dispersion profile to the data, we obtain constraints on the concentration 
parameter and the slope of the mass distribution (the right panel of Supplementary Fig.~2). For simplicity, 
we assume that the concentration parameter does not vary with the virial mass. The obtained 
concentrations are smaller by $20-30$ per cent than those found for simulated dark matter halos\cite{Kly10}. 
This effective flattening of the mass profile most (smaller concentration parameters) 
arises from the fact that the gravitational potential of clusters is measured with respect to BCGs 
whose positions exhibit random off-sets from the true cluster centres\cite{Lin07}.

We make use of the constraints on $c_{\rm v}$ and $\alpha$ to calculate the profile of gravitational redshift 
given by eq. (\ref{shift-stack}), where the gravitational potential takes the NFW form, i.e.
\begin{equation}\label{psiNFW}
\Phi(r)=-(GM_{\rm v}/r_{\rm v})^{1/2}g(c_{\rm v})^{1/2}\frac{\ln(1+r/r_{\rm v})}{r/r_{\rm v}}
\end{equation}
and $g(c_{\rm v})=1/[\ln(1+c_{\rm v})-c_{\rm v}/(1+c_{\rm v})]$, and the number density of galaxies 
is proportional to that of dark matter. The resulting $\Delta$ profiles are shown in Supplementary Fig.~3. The 
blue (red) profile corresponds to an isotropic (anisotropic) model of galaxy orbits and the widths of the profiles 
are the $1\sigma$ ranges obtained by marginalising over all free parameters of the model. Both profiles are fully 
consistent with the constraints on $\Delta$ inferred from the observed velocity distributions (black points). 
Since the systematic errors induced by a choice of the anisotropy model are much smaller than the random errors 
associated with the gravitational redshift measurement, in the final comparison we only consider the profile 
calculated for an anisotropic model of galaxy orbits-- a more reliable model in recovering the kinematics of galaxy 
clusters\cite{Woj10}. We also note the gravitational potential resulting from the analysis with anisotropic galaxy 
orbits is the basis for the calculations of the gravitational redshift effect in alternative models of gravity (see the 
last section of SI).

The theoretical calculations presented in this section rely on the extrapolation of the NFW profile beyond the virial 
sphere which is still robust to $2r_{\rm v}$, but is probably less justified for larger radii\cite{Tav08}. In order to check 
the potential impact of this assumption on the estimation of $\Delta(R)$, we considered a set of simple modifications of 
the density profile at large radii. We found that varying the asymptotic slopes of the dark matter (or galaxy number) 
density profile by $\pm 0.5$ at $r>2r_{\rm v}$ does not change the $\Delta$ profiles by more than $4$ per cent. 
We also found that the mass limits imposed on the mass distribution have a negligible effect on the final 
$\Delta$ profile. We verified that changing both mass limits by $50$ per cent induces error which do not 
exceed the uncertainty due to the unknown anisotropy of galaxy orbits.

\section*{{\large Test on a mock data sample}}

In order to test the statistical robustness of the gravitational redshift detection, we analyse 
mock kinematic data generated from cosmological $N$-body simulations of a standard $\Lambda$CDM 
cosmological model (for details of the simulations see Wojtak {\it et al.}\cite{Woj08}). Such test allows 
to check whether all effects related to the internal substructure of clusters and their perturbed velocity 
distributions are sufficiently reduced in the procedure of stacking kinematic data of a large number 
of galaxy clusters.

We generate mock redshift data by drawing randomly dark matter particles from $7,800$ cylinders 
of observation (the number of cylinders corresponds to the number of clusters in the real data sample compiled 
from the SDSS). The cylinders of observations are defined by the $\pm 4000$ km s$^{-1}$ velocity range and 
the $6$ Mpc aperture. Their viewing angles are chosen at random and geometrical centres are located 
at the centre of one of $80$ cluster-mass dark matter halos. In order to simulate the offset between the positions 
of the cluster central galaxies and the cluster mass centres, we introduce a random shift 
between the positions and velocities of the cylinders and the corresponding halo mass centres. 
The maximum value of this shift is $0.1$ $h^{-1}$ Mpc for the positions\cite{Lin07} and $\pm35$ per cent of the total 
velocity dispersion for velocities 
projected onto the line of sight\cite{Ski11}. The total number of velocities in a composite cluster is fixed at $10^{5}$ 
or $10^{6}$. The former corresponds to the number of redshifts in the current SDSS sample and the latter 
represents a forecast for the future.

In order to perform a test of detectability of the gravitational redshift effect, we consider two data samples. 
The first consists of velocities which are not corrected for the gravitational redshift and the second 
takes into account an additional velocity shift proportional to the local gravitational potential. In both cases 
particle positions remain the same. Both mock data samples are analysed in the same manner as the SDSS data. The 
obtained constraints on the mean velocity as a function of the projected radius are shown in Supplementary Fig.~4, 
where blue and red colour refer to the samples with $10^{5}$ and $10^{6}$ redshifts, respectively.

The mean velocity obtained for the sample without gravitational redshift 
(points with the dashed error bars) are consistent with $\Delta=0$ at all radii. This shows that the number of 
clusters used in our work is sufficient to reduce all local effects giving rise to the fluctuations of the mean 
velocity. Velocity shifts obtained for the second data sample (points with the solid error bars) clearly indicate 
the presence of the gravitational redshift effect. They also trace the true profile of this effect 
averaged over the halo sample used for generating mock kinematic data (black solid line). We find that the 
measured velocity shifts for $10^{5}$ data points (blue points with the solid error bars) deviate from $\Delta=0$ 
profile at nearly the same significance level as the results obtained for the SDSS data. We also note that an 
increase of the number of redshifts per cluster improves the errors of the measurement according to the rule of 
the inverse square root proportionality.

\section*{{\large Gravitational redshift in alternative models of gravity}}

The only difference in the calculation of the gravitational redshift for models of modified gravity relies on 
replacing the Newtonian gravitational potential in eq.~(\ref{vel-shift-proj}) by the potential emerging from 
a given gravity. For the $f(R)$ model, the gravitational potential is calculated using phenomenological relations 
obtained by Schmidt\cite{Sch10} who ran a series of cosmological simulations of this model and quantified 
its effect on cluster dynamics in terms of the effective enhancement of gravitational acceleration with respect to 
the standard gravity. In our work we consider the most critical case of his results 
(with $|\textrm{d}f/\textrm{d}R|\approx10^{-4}$) leading to a homogeneous amplification of gravitational acceleration 
at all radii by factor of $1.33$.

Gravitational potential in TeVeS theory is a sum of the Newtonian potential $\Phi_{\rm N}$ and the scalar field 
$\phi$. The former is calculated assuming $80$ per cent of the total mass inferred from the velocity dispersion profile 
in the framework of Newtonian gravity. This factor lowers the total-to-barynic mass ratio to the value estimated in 
galaxy clusters under assumption of the Modified Newtonian Dynamics\cite{Poi05}. Assuming spherical symmetry, the scalar 
field $\phi$ is a solution of the following equation\cite{Bek04}
\begin{equation}\label{scalar}
\mu(y)\nabla\phi=(k/4\pi)\nabla\Phi_{\rm N},
\end{equation}
where $y=(\nabla\phi)^2kl^{2}$, $k<<1$ and $\sqrt{k}/(4\pi l)\approx a_{0}=10^{-10}$ m s$^{-2}$. In our calculations 
we adopted $\mu(y)=\sqrt{y}/(1+\sqrt{y})$, which is one of the commonly used interpolating function in Modified 
Newtonian Dynamics\cite{Mil83}, and $k=0.01$.

In both schemes of the calculation, the reference gravitational potential of Newtonian gravity is given by the 
constraints from fitting the velocity dispersion profile with an anisotropic model of galaxy orbits. We note that using the results 
of an isotropic model does not change the main conclusion of the test summarised in Fig.~3: the $f(R)$ model 
is still consistent with the data, whereas TeVeS yields even more divergent profile of the gravitational redshift.

\newpage

\begin{center}
    \leavevmode
    \epsfxsize=16cm
    \epsfbox[50 50 580 420]{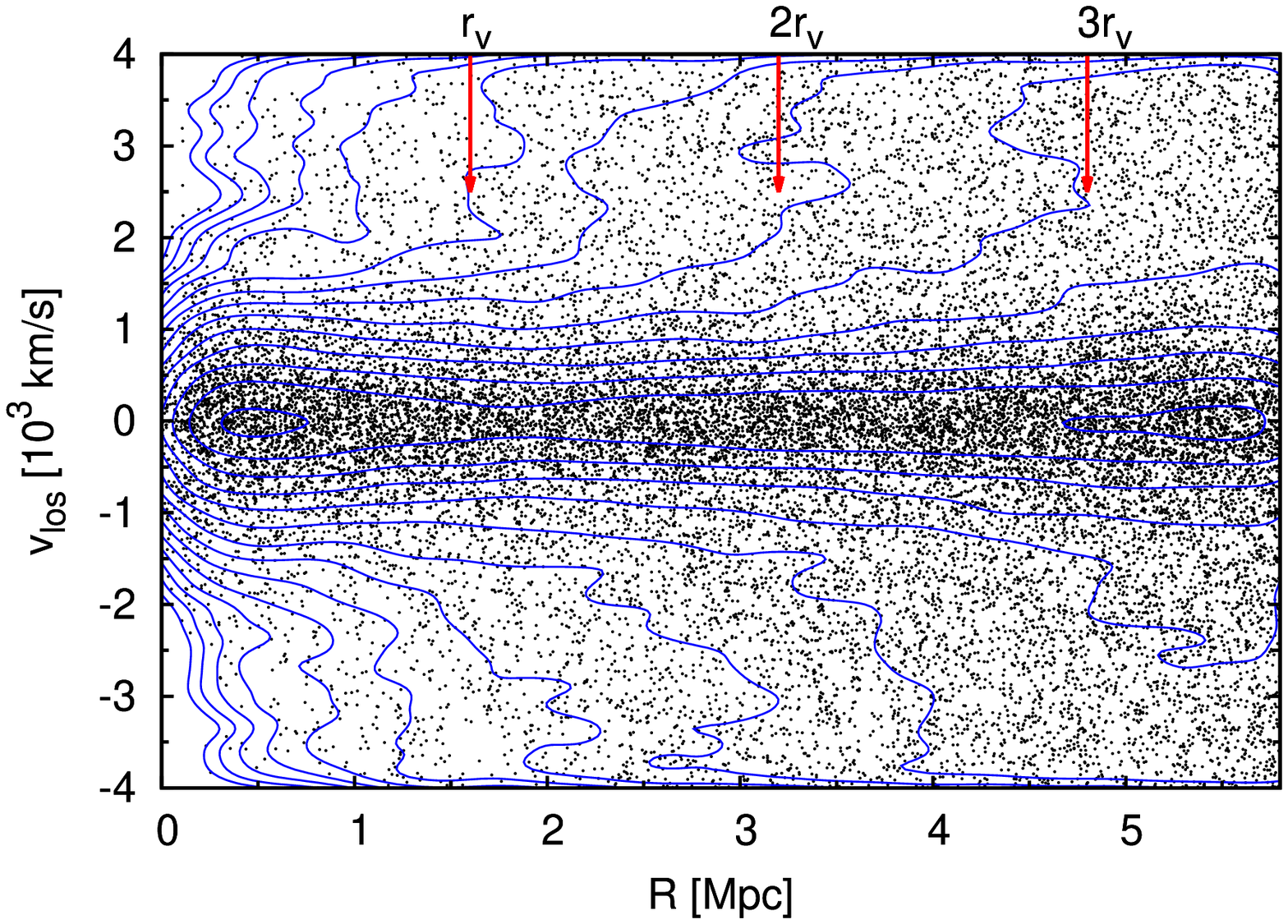}
\end{center}
\vspace{-0.5cm}
{\sf \textbf{Supplementary Figure~1} Velocity diagram combined from kinematic data of $7800$ galaxy clusters 
detected in the SDSS\cite{Hao10} Data Release 7. Velocities $v_{\rm los}$ of galaxies with 
respect to the brightest cluster galaxies are plotted as a function of the projected cluster-centric 
distance $R$. Blue lines are the iso-density contours equally spaced in the logarithm of galaxy 
density in the $v_{\rm los}-R$ plane. The arrows show characteristic scales related to the mean 
virial radius estimated in dynamical analysis of the velocity dispersion profile. Data points 
represent $20$ per cent of the total sample.}

\newpage

\begin{center}
    \leavevmode
    \epsfxsize=16cm
    \epsfbox[50 50 1110 420]{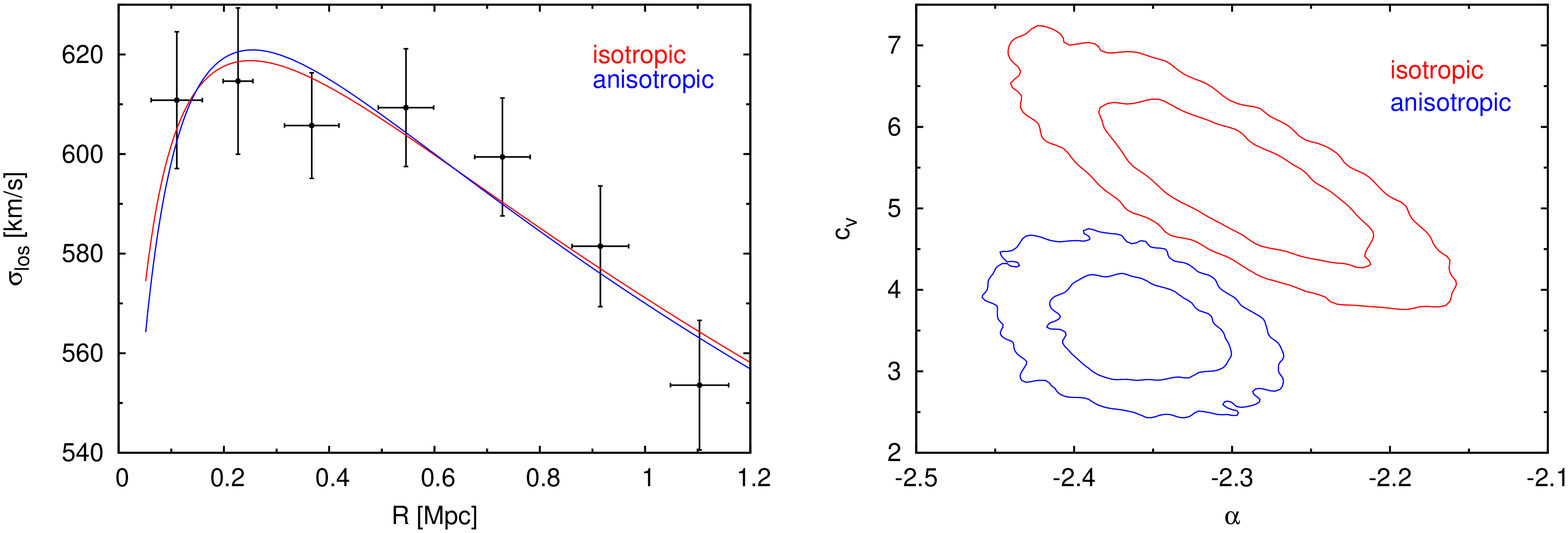}
\end{center}
{\sf \textbf{Supplementary Figure~2} Velocity dispersion profile of the composite cluster (left panel) and constraints 
on the concentration parameter $c_{\rm v}$ and the logarithmic slope of the mass distribution $\alpha$ 
(right panel) from fitting the velocity dispersion profile with an isotropic (blue) or anisotropic 
(red) model of galaxy orbits. The solid lines in the left panel show the best-fitting profiles of the velocity 
dispersion profile. The contours in the right panel are the boundaries of the $1\sigma$ and $2\sigma$ 
confidence regions of the likelihood function. The error bars in the left panel represent the range containing 
$68$ per cent of the marginal probability.}

\newpage

\begin{center}
    \leavevmode
    \epsfxsize=16cm
    \epsfbox[50 50 580 420]{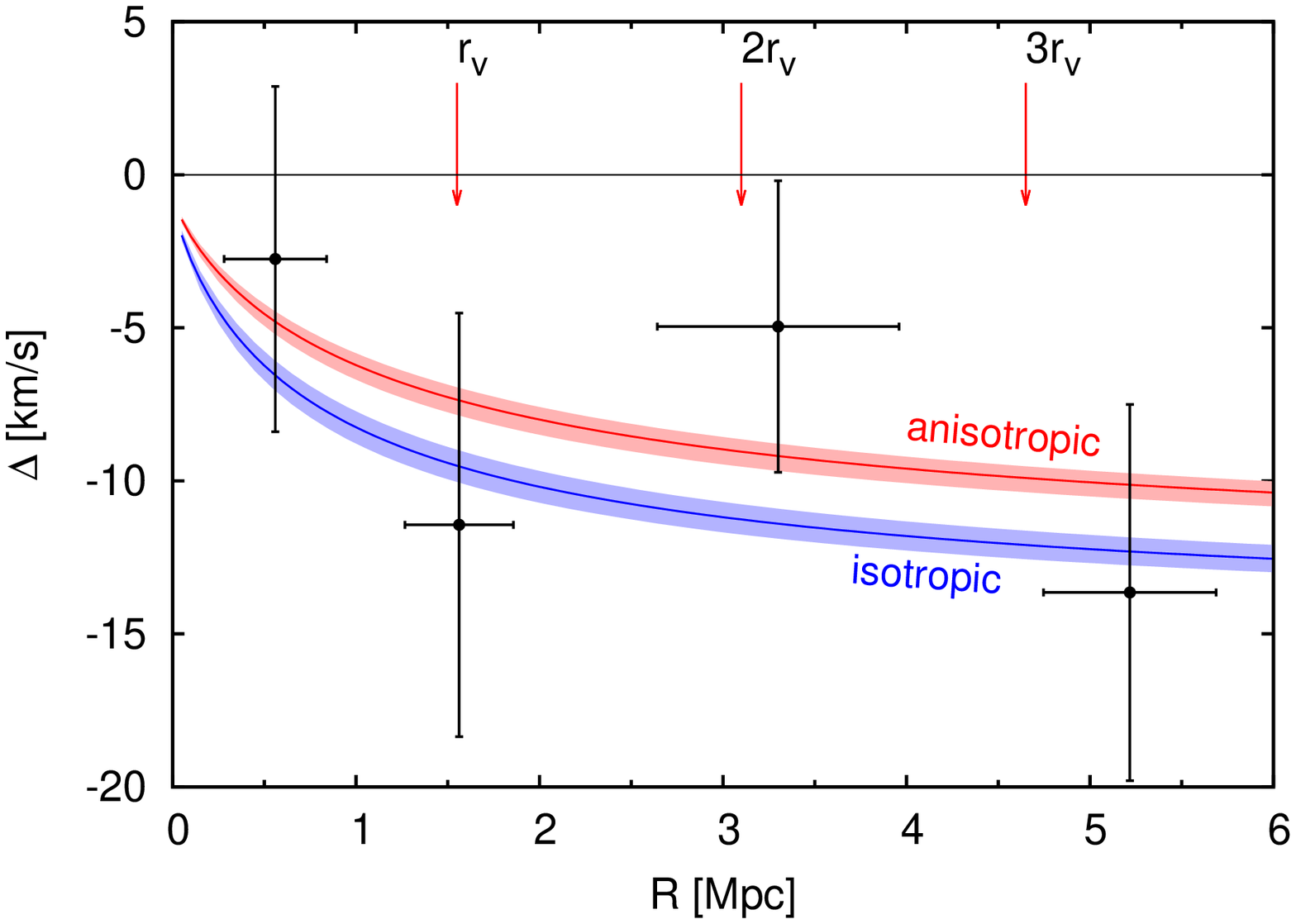}
\end{center}
\vspace{-0.5cm}
{\sf \textbf{Supplementary Figure~3} Theoretical predictions for gravitational redshift in terms of the mean velocity $\Delta$. 
The blue and red profiles were calculated on the basis of the mean cluster gravitational potential inferred 
from fitting the velocity dispersion profile with an isotropic  and anisotropic model of galaxy orbits, respectively 
The widths of both profiles represent the range of $\Delta$ containing $68$ per cent of the probability marginalised over parameters 
of the gravitational potential and the distribution of clusters masses in the sample. 
The arrows show characteristic scales related to the mean virial radius $r_{\rm v}$. Black points show constraints on 
gravitational redshift from the analysis of the observed velocity distributions. The error bars represent the range of 
$\Delta$ parameter containing $68$ per cent of the marginal probability and the dispersion of the projected radii in 
a given bin.}

\newpage

\begin{center}
    \leavevmode
    \epsfxsize=16cm
    \epsfbox[50 50 580 420]{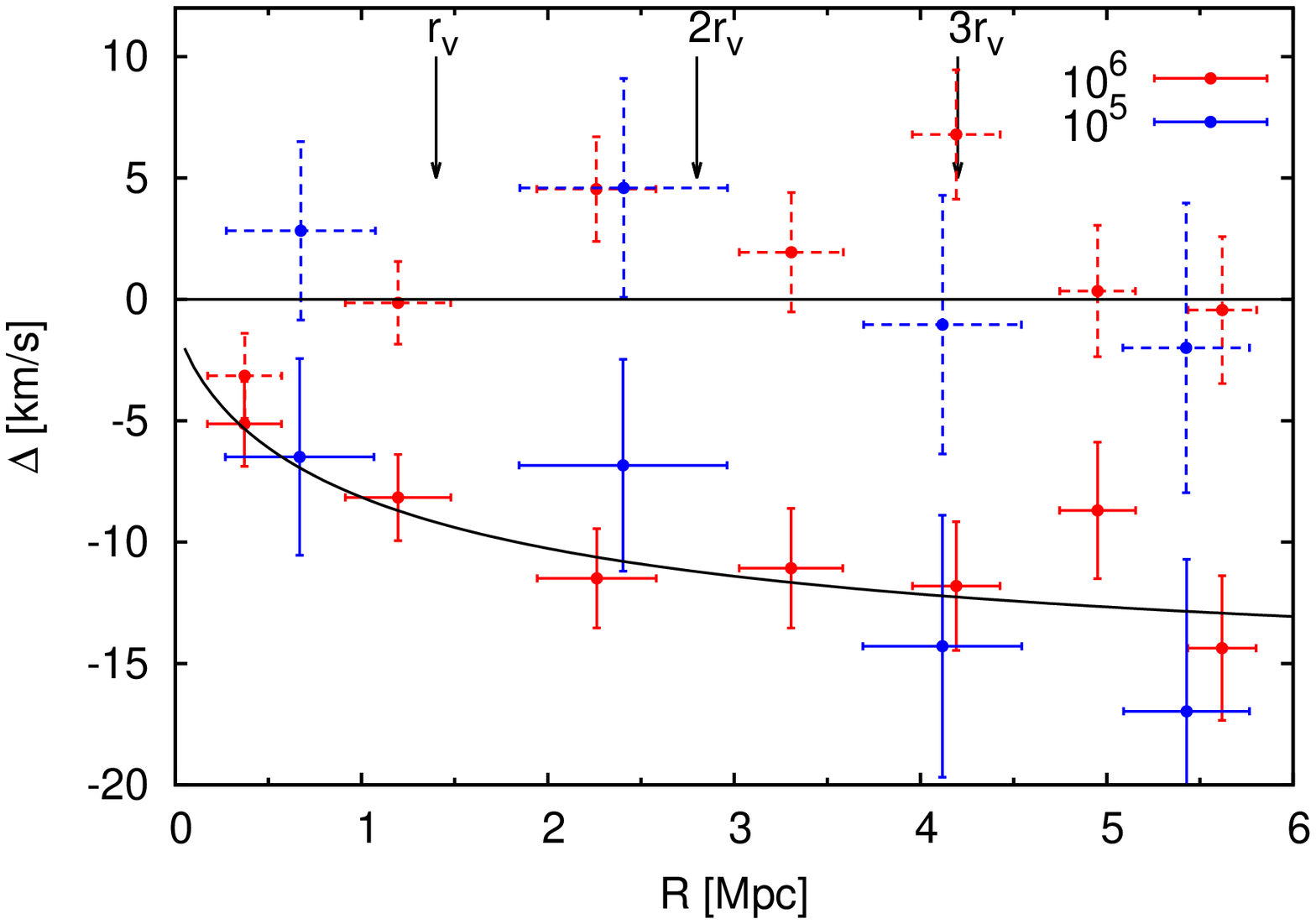}
\end{center}
\vspace{-0.5cm}
{\sf \textbf{Supplementary Figure~4} Constraints on the velocity shift in mock composite clusters generated from cosmological $N$-body 
simulations. Blue and red points correspond to the data samples with $10^{5}$ and $10^{6}$ redshifts, respectively. 
Data points with the solid (dashed) error bars represent the results obtained for the samples which include (do not 
include) the effect of gravitational redshift. Black solid line shows the true profile of the velocity shift $\Delta$ 
caused by gravitational redshift. The error bars represent the range of $\Delta$ parameter containing $68$ per cent of 
the marginal probability and the dispersion of the projected radii in a given bin. The arrows show characteristic scales 
related to the mean virial radius $r_{\rm v}$.}

\end{document}